\documentclass[letterpaper,12pt]{JHEP3}
\usepackage{bm}
\usepackage{amsmath}
\usepackage{yfonts}
\usepackage{epsfig}
\usepackage{psfrag}
\usepackage{cite}
\usepackage{verbatim}
\newcommand{\qn}{\textswab{q}}
\newcommand{\wn}{\textswab{w}}
\newcommand{\mn}{\textswab{m}}
\newcommand{\<}{\langle}
\renewcommand{\>}{\rangle}
\renewcommand{\d}{\partial}

\newcommand{\q}{\bm{q}}
\newcommand{\x}{\bm{x}}

\renewcommand{\Im}{\mathrm{Im}\,}

\newcommand{\stru}{\rule[-.2in]{0in}{.2in}}

\def\calv         {{\cal V}}

\def\reals        {{\mathbb R}}

\def\del          {\partial}

\def\ee           {{\rm e}}

\def\Im           {{\rm Im\hskip0.1em}}

\def\sqr#1#2{{\vcenter{\vbox{\hrule height.#2pt  
 \hbox{\vrule width.#2pt height#1pt \kern#1pt
 \vrule width.#2pt}\hrule height.#2pt}}}}

\newcommand{\ft}[2]{{\textstyle{\frac{#1}{#2}}}}
\def\jsquare{\mathop{\mathchoice{\sqr{8}{32}}{\sqr{9}{12}}
{\sqr{6.3}{9}}{\sqr{4.5}{9}}}}

\def\b{\beta}

\def\rp{r_+}

\def\be{\begin{equation}}
\def\ee{\end{equation}}

\long\def\eqlabel#1{\ifnum\draftcontrol=1
                    \tag@false  
                    \tag*{(\theequation) \hbox to -0.2cm{\hspace{0cm}\small{#1}\hss}}
                    \refstepcounter{equation} 
                    \edef\@currentlabel{\theequation}
                    \ltx@label{#1}          
                    \else
                    \label{#1}
                    \fi
                    }

\title{Hydrodynamics of R-charged black holes}

\author{Dam T.~Son\\
Institute for Nuclear Theory, University of Washington,
Seattle, WA 98195, USA\\
Emails: \email{son@phys.washington.edu}
}
\author{ Andrei O.~Starinets\\
Perimeter Institute for Theoretical Physics,
Waterloo, ON N2L 2Y5, Canada\\
E-mail: \email{starina@perimeterinstitute.ca}
}

\date{April 2004}

\abstract{
We consider hydrodynamics of ${\cal N}=4$ supersymmetric 
$SU(N_c)$ Yang-Mills plasma at a nonzero density of $R$-charge.
In the regime of large $N_c$ and large 't Hooft coupling 
the gravity dual description involves an  asymptotically Anti- de Sitter 
five-dimensional 
charged black hole solution of Behrnd, Cveti\v{c} and Sabra.
We compute the shear viscosity as a function of 
chemical potentials conjugated to the three $U(1) \subset SO(6)_R$
 charges. The ratio of the 
 shear viscosity to entropy density is independent of 
the chemical potentials and is equal to $1/4\pi$. For a 
single charge black hole we also compute the thermal conductivity, 
and investigate the critical behavior of the transport coefficients near 
the boundary of thermodynamic stability. 
}

\begin{document}
\section{Introduction}

Recent studies of strongly coupled thermal gauge theories in the framework of
 the gauge-gravity duality (\cite{Maldacena:1997re}, \cite{Gubser:1998bc},
 \cite{Witten:1998qj}, for a review see 
\cite{Aharony:1999ti}) 
 suggest that in all those theories 
in the regime described by 
gravity duals the ratio of the shear viscosity to volume 
entropy density is universal and equal to $1/4\pi$ \cite{Kovtun:2003wp},
 \cite{Kovtun:2004de}, \cite{Buchel:2004qq}.
 As this result was obtained assuming 
zero densities of conserved charges, a natural question to ask is 
what happens when the chemical
potentials conjugated to these charges are turned on. 

The simplest ten-dimensional gravitational background corresponding to a non-zero chemical
 potential in a dual four-dimensional finite-temperature field theory is the one of spinning
 near-extremal three-branes \cite{Gubser:1998jb}, \cite{Chamblin:1999tk}, 
\cite{Cvetic:1999ne}.  
The number of independent commuting angular momenta that can be given to the three-branes
is equal to the rank $r=3$ of the isometry group $SO(6)$ of the space transverse to the branes.  
Upon dimensional reduction on $S^5$ one obtains the background corresponding to the five-dimensional asymptotically AdS black hole of the Reissner-Nordstr\"om type with three 
$U(1)$ charges proportional to the angular momenta of the branes \cite{Cvetic:1999ne}.
 This background was found by 
  Behrnd, Cveti\v{c} and Sabra \cite{Behrndt:1998jd} 
as a particular solution to the equations of motion of the
 five-dimensional ${\cal N}=2$ gauged supergravity. The solution corresponding to the 
dimensional reduction of the spinning three-brane metric has a translationally invariant horizon.
(It can also be regarded as a black hole with a spherical horizon in an infinite volume limit.)

In the AdS/CFT correspondence the isometry group of $S^5$ is interpreted as the $R$-symmetry group 
of the dual ${\cal N}=4$ supersymmetric Yang-Mills (SYM) theory. 
Three independent chemical potentials $\mu_i$ can be 
 introduced as the Lagrange multipliers to the three $U(1)$ charges in the Cartan subalgebra of $SO(6)_R$. 
Thermodynamics of the R-charged black holes in the context of the AdS/CFT 
correspondence was originally studied in  \cite{Gubser:1998jb}, \cite{Cai:1998ji}, 
\cite{Chamblin:1999tk}, 
\cite{Cvetic:1999ne}. One feature relevant for our discussion
 is the thermodynamic instability\footnote{The related instability of the 
gravitational background was studied in \cite{Gubser:2000ec}.}
occurring for black holes with excessively large charge 
(or equivalently for the three-branes rotating too fast)  \cite{Gubser:1998jb}, 
\cite{Cai:1998ji}. The instability 
may signal the onset of a phase transition. However, both in
 gravity and in the dual strongly coupled field theory picture it
 is not clear what the
 new phase might be.  

In this paper we study the hydrodynamic regime of the four-dimensional 
 ${\cal N}=4$ $SU(N_c)$ SYM theory with 
three non-zero chemical potentials in the limit of large $N_c$ and 
large 't Hooft coupling 
$g^2_{YM} N_c$. Using Kubo formula, 
we compute the shear viscosity as a function of the three charges 
(or chemical potentials) and show that for any values of the charges in the 
thermodynamic stability domain the ratio of the viscosity to entropy density is
 equal to $1/4\pi$. For a technically simpler case of a single charge black hole we
explicitly compute thermal correlation functions of the stress-energy tensors and R-currents
 in the shear channel in the hydrodynamic approximation. The correlators exhibit a 
diffusion pole with the dispersion relation that confirms the result for the shear
 viscosity found from the Kubo formula. We also compute thermal conductivity. 
The ratio of thermal conductivity and shear viscosity obeys a simple relation reminiscent
 of the Wiedemann-Franz law for the 
 ratio of the thermal conductivity to the electrical
 conductivity.
Finally, we investigate the behavior of the transport coefficients
 near the boundary of thermodynamic stability and compute the corresponding critical exponent.

The paper is organized as follows. We review the  STU-model solution of 
Behrnd, Cveti\v{c} and Sabra in Section \ref{stu}. 
The shear viscosity for the three-charge black hole 
solutions is computed in Section \ref{kubo_shear}. Starting from Section \ref{shear}, 
we specialize 
to the single-charge background: in Section \ref{shear} we compute the correlation functions of
the shear mode components of  
the stress-energy tensors and R-currents in the hydrodynamic approximation, in Section \ref{conductivity}  we 
obtain the thermal conductivity, and in Section \ref{critical}
 we discuss the critical behavior 
of the transport coefficients. 
For convenience of the reader, in Appendix \ref{appendix_A}
 we outline the rescaling procedure necessary to 
obtain an asymptotically AdS charged black hole with a translationally invariant horizon from the
 black hole with a spherical horizon.
In Appendix \ref{kubo_tc} we review relativistic hydrodynamics at finite 
chemical potential.

\section{The R-charged black hole background}
\label{stu}

The STU-model solution to equations of 
motion of $D=5$ ${\cal N}=2$ gauged supergravity was found by
Behrnd, Cveti\v{c} and Sabra \cite{Behrndt:1998jd}. The relevant part of the gauged supergravity
effective Lagrangian is 
\begin{equation}
 {{\cal L}\over
 \sqrt{-g} }=
 R + {2\over L^2} {\cal V} - {1\over 2} G_{ij} F_{\mu\nu}^i 
 F^{\mu \nu\, j} -G_{ij} \partial_{\mu} X^i \partial^{\mu} X^j
+ {1\over 24  \sqrt{-g} } \epsilon^{\mu\nu\rho\sigma\lambda}\epsilon_{ijk} 
 F_{\mu\nu}^i F^{\rho \sigma j} A_\lambda^k\,,
\label{lagrangian}
\end{equation}
where $F_{\mu\nu}^i$, $i=1,2,3$ are the field-strength tensors of the three 
Abelian gauge fields, $X^i$ are three real scalar fields subject 
to the constraint $X^1 X^2 X^3=1$. The metric on the scalar manifold
 is given by
$$
G_{ij} = {1\over 2} \mbox{diag} \left[ (X^1)^{-2}, \, (X^1)^{-2}, \,(X^1)^{-2}
\right]\,.
$$
The scalar potential is
$$
 {\cal V} = 2 \sum\limits_{i=1}^3 {1\over X^i}\,.
$$
The three-charge non-extremal STU solution is specified by
the following background values of the metric
\begin{equation}
ds^2 = - {\cal H}^{-2/3}\, f_k \, dt^2 
+ {\cal H}^{1/3} \left( f_k^{-1} dr^2 + r^2 d \Omega_{3,k}^2\right)\,,
\label{metric_k}
\end{equation}
%
\begin{equation}
f_k = k - {m_k\over r^2} + {r^2\over L^2}{\cal H} \,, \qquad 
 H_i = 1 + {q_i\over r^2}\,, \qquad  {\cal H} = H_1 H_2 H_3 \,,
\end{equation}
 as well as the scalar and the gauge fields
\begin{equation}
X^i = {{\cal H}^{1/3}\over H_i} \,,  \qquad
A^i_t = \sqrt{ {k q_i + m_k\over q_i}} \left( 1 - H_i^{-1}\right)\,. 
\end{equation}
The parameter $k$ determines the spatial curvature of  $d \Omega_{3,k}^2$:
$k=1$ corresponds to the metric on the three-sphere of unit radius,
$k=0$ - to the metric on $\reals^3$.
It was shown in \cite{Cvetic:1999ne} that the $k=0$ solution arises as the
 Kaluza-Klein reduction on $S^5$ of the ten-dimensional metric 
describing spinning near-extremal three-branes. The three R-charges
 $q_i$ are related to the three independent angular momenta in ten 
 dimensions.
Since, strictly speaking, hydrodynamic regime is meaningful only in the 
case of a translationally-invariant horizon, in this paper we set\footnote{The background with a translationally-invariant horizon and related thermodynamics can also be  obtained by taking an infinite volume limit of the $k=1$ solution (see Appendix  \ref{appendix_A}).}
$k=0$
and
$$
d \Omega_{3,0}^2 \rightarrow {1\over L^2} \left( dx^2 + dy^2 + dz^2\right)\,.
$$
Introducing the new radial coordinate $u=r_+^2/r^2$, 
where $r_+$ is the largest root of the equation $f(r)=0$, 
we write the background fields in the form
\begin{equation}
ds^2_5 = - {\cal H}^{-2/3}{(\pi T_0 L)^2 \over u}\,f \, dt^2 
+  {\cal H}^{1/3}{(\pi T_0 L)^2 \over u}\, \left( dx^2 + dy^2 + dz^2\right)
+ {\cal H}^{1/3}{L^2 \over 4 f u^2} du^2\,,
\label{metric_u_3}
\end{equation}
\begin{equation}
f(u) = {\cal H} (u) - u^2 \prod\limits_{i=1}^3 (1+\kappa_i)\,, 
\;\;\;\;\; H_i = 1 + \kappa_i u \,, \;\;\;\;\; 
\kappa_i \equiv {q_i\over r_+^2}\,, \;\;\;\;\; T_0 = r_+/\pi L^2\,.
\label{identif}
\end{equation}
The scalar fields and the 
gauge fields are given by\footnote{We change the normalization of 
the gauge fields by a factor of $\sqrt{2}/L$. 
This normalization is used in the rest of the paper.}
\begin{equation}
X^i = {{\cal H}^{1/3}\over H_i(u)} \,, \qquad 
A^i_t = {\tilde{\kappa}_i \sqrt{2} u\over L H_i(u)}\,,\qquad  
\tilde{\kappa}_i = {\sqrt{q_i}\over L} 
 \prod\limits_{i=1}^3 (1+\kappa_i)^{1/2}
\,. 
\label{scal_gauge_u_3}
\end{equation}

The Hawking temperature of the background 
(\ref{metric_u_3}) is given by
\begin{equation}
T_H = 
{2 + \kappa_1 + \kappa_2 + \kappa_3 - \kappa_1 \kappa_2  \kappa_3\over 
2\sqrt{(1+\kappa_1)(1+\kappa_2) (1+\kappa_3)}}\, T_0\,.
\end{equation}
The volume density of the Bekenstein-Hawking entropy is 
\begin{equation}
s = {A_H\over 4 G_5 V_3} = {\pi^2 N^2 T_0^3 \over 2} 
 \prod\limits_{i=1}^3 (1+\kappa_i)^{1/2}  \, , 
\label{entropy_density}
\end{equation}
where $V_3$ is the spatial volume along the three infinite dimensions 
of the horizon, $G_5=\pi L^3/2 N^2$. 
The energy density and pressure are given by
\begin{eqnarray}
\varepsilon &=& { 3 \pi^2 N^2 T_0^4 \over 8}   \prod\limits_{i=1}^3
 (1+\kappa_i)\,,\label{energy_density} \\
  P &=&  {\pi^2 N^2 T_0^4 \over 8}   \prod\limits_{i=1}^3
 (1+\kappa_i)\,. 
\label{pressure}
\end{eqnarray}
The densities of physical charges are 
\begin{equation}
\rho_i = {\pi \over 4 G_5} {r_+^2 \tilde{\kappa}_i L\over \sqrt{2} V_3} =
{\pi N^2 T_0^3\over 8} \sqrt{2 \kappa_i}  \prod\limits_{l=1}^3
 (1+\kappa_l)^{1/2}\,.  
\end{equation}
The chemical potentials conjugated to $\rho_i$ are defined as
\begin{equation}
\mu_i = A_t^i (u)\Biggl|_{u=1} = {\pi T_0 \sqrt{2 \kappa_i}\over (1+\kappa_i)}
 \prod\limits_{l=1}^3
 (1+\kappa_l)^{1/2}\,.  
\end{equation}
For the grand canonical ensemble, 
where the system in thermodynamic equilibrium  
is characterized by the values of temperature and chemical potentials
regulating its interaction with the surrounding heat bath, 
the appropriate thermodynamic potential is the Gibbs potential 
 $\Omega$,
\begin{equation}
\Omega /V_3 = - P = \varepsilon - T_H s - \sum\limits_{i=1}^3 \mu_i \rho_i \,.
\end{equation}
The first law of thermodynamics reads
\begin{equation}
d P = s d T_H +  \sum\limits_{i=1}^3  \rho_i d \mu_i\,.
\end{equation}
A stable thermodynamic equilibrium is determined by the conditions
\begin{equation}
 \left( \delta \Omega
 \right)_{T,\mu_i
 \; \mbox{\tiny{fixed}}} = 0\,, \qquad  \;\;\;\; \left( \delta^2 \Omega
 \right)_{T,\mu_i
 \; \mbox{\tiny{fixed}}} > 0\,.
\label{stability}
\end{equation}
The stability condition (\ref{stability}) translates into the equation
\begin{equation}
 \det \; \Biggl(
{\partial^2 \varepsilon (s,\rho_i) \over \partial s \partial \rho_i } \Biggr)> 0\,.
\end{equation}
Since $\kappa_i = 8 \pi^2 \rho_i^2/s^2$ and
$$
\varepsilon (s,\rho_i) = {3 s^{4/3}\over 2 (2\pi N)^{2/3}} \, 
 \prod\limits_{i=1}^3 \left( 1 + {8\pi^2 \rho_i^2\over s^2}\right)^{1/3}\,,
$$
 the condition of thermodynamic stability implies 
the following  constraint on $\kappa_i$
\begin{equation}
2 - \kappa_1 -  \kappa_2  - \kappa_3 + \kappa_1\,  \kappa_2\,   \kappa_3 > 0\,.
\end{equation}

\section{Shear viscosity}
\label{kubo_shear}

The simplest way to compute shear viscosity
from the dual gravity background is to use Kubo formula.
Kubo formula  relates
the shear viscosity to
the correlation function of the stress-energy tensor at zero 
spatial momentum,
\begin{equation}\label{kubo_visc}
\eta = \lim_{\omega \rightarrow 0} {1\over 2\omega} 
 \int\!dt\,d\x\, e^{i\omega t}\,
\langle [ T_{xy}(x),\,  T_{xy}(0)]\rangle = 
- \lim_{\omega \rightarrow 0} {\mbox{Im} 
G (\omega,0)\over \omega}\,,
\end{equation}
where  the
retarded Green's function for the components of the stress-energy tensor
is defined as
\begin{equation}
  G_{\mu\nu\lambda\rho} (\omega, \q)
  = -i\!\int\!d^4x\,e^{-iq\cdot x}\,
  \theta(t) \< [T_{\mu\nu}(x),\, T_{\lambda\rho}(0)] \>\,.
\end{equation}
Thus finding the shear viscosity amounts to 
computing the zero-frequency limit of the imaginary part of the 
retarded correlator $G_{xy xy}(\omega, q)$. To compute the correlator, we follow the
procedure outlined in \cite{Son:2002sd} and used in \cite{Policastro:2002se}
to determine the shear viscosity of the strongly coupled ${\cal N}=4$ SYM 
at zero chemical potential.
First, one has to determine the equation obeyed by the component
 $h_{xy}(u,t,z)$
of the gravitational perturbation of the background
 (\ref{metric_u_3}) - (\ref{scal_gauge_u_3}). By symmetry argument 
 \cite{Policastro:2002se},
 \cite{Kovtun:2005ev}
or by 
the direct analysis of perturbations of the equations
 of motion 
following from the Lagrangian (\ref{lagrangian}) one can show that
the Fourier component of the off-diagonal perturbation  $\phi \equiv h^x_y$
decouples from  all
 other perturbations and  obeys the equation
for a minimally coupled massless scalar in the background (\ref{metric_u_3})
\begin{equation}
\phi_k'' + {u f' -f\over u f}\, \phi_k' + 
{{\cal H}\, \wn^2 - f\, \qn^2\over u f^2}\, \phi_k =0\,,
\label{scal_3}
\end{equation}
where
\begin{equation}
\wn = {\omega \over 2 \pi T_0}\,, \;\; \qquad \; \; 
\qn = {\omega \over 2 \pi T_0}\,.
\end{equation}
The solution to Eq.~(\ref{scal_3}) in the hydrodynamic regime
 $\wn \ll 1$, $\qn\ll1 $ can be obtained along the lines of 
Ref.~\cite{Policastro:2002se}. We find
\begin{eqnarray}
\phi_k &=& C_k f^{- i\, \wn\, U}
\Biggl\{ 1 - {i U\over 2} 
\Biggl[ \log{ {c^3 (u-1)^2\over a(u-1)^2 + b (u-1) +c  } } \nonumber \\
&-&
 {b-2c\over \sqrt{b^2-4 a c}} 
\log{  {(2 a (u-1) + b -  \sqrt{b^2-4 a c})(b+
   \sqrt{b^2-4 a c})\over (2 a (u-1) + b +  \sqrt{b^2-4 a c})(b-
   \sqrt{b^2-4 a c})}   }\Biggl] +\cdots \Biggr\}\,,
\label{scal_sol_3}
\end{eqnarray}
where ellipses denote higher order terms in the hydrodynamic expansion,
$$
a = \prod\limits_{i=1}^3 \kappa_i\,,\qquad
b = 2 a  -  1 -
\sum\limits_{i=1}^3 \kappa_i\,,\qquad c = b-a-1\,.
$$
and 
$$
U(\kappa_1, \kappa_2, \kappa_3)  =  {\sqrt{(1+\kappa_1)(1+\kappa_2) 
(1+\kappa_3)}\over 2 + \kappa_1 +\kappa_2 +\kappa_3 - 
\kappa_1\kappa_2\kappa_2  } \,.
$$
In the limit $\kappa_i \rightarrow 0$ the solution (\ref{scal_sol_3})
reduces to the one found in  \cite{Policastro:2002se}.

Another essential ingredient in computing the correlator is the 
boundary action. The total action is given by
\begin{equation}
S = {1\over 16 \pi G_5} \int_{{\cal M}_5}\, d^5 x \; {\cal L} + 
 {1\over 8 \pi G_5} \int_{\partial {\cal M}_5}\, d^4 x \; \sqrt{-h} \, K +
 {1\over 8 \pi G_5} \int_{\partial {\cal M}_5}\, d^4 x  \sqrt{-h}\, W\,,
\label{total_action_3}
\end{equation}
where the second term is the Gibbons-Hawking boundary term,
 and the third term is required to make the action finite\footnote{
For a discussion of relevant issues in the context of 
  holographic renormalization, see e.g. \cite{Balasubramanian:1999re}, 
 \cite{Bianchi:2001kw}, \cite{Skenderis:2002wp}, \cite{Buchel:2003re}, \cite{Batrachenko:2004fd}.}
 in the limit
$u\rightarrow 0$.
The explicit form of $W$ for the background 
of interest was determined in \cite{Batrachenko:2004fd}:
\begin{equation}
W = -{  {\cal H}^{1/3} \over L} \, \sum\limits_{i=1}^3 H_i^{-1} \,.
\end{equation}
Computing the action (\ref{total_action_3}) on shell
and expanding to quadratic order in  $h_x^y$ we find the following 
boundary action for $\phi_k$
\begin{equation}
S_B = - {\pi^2 N^2 T_0^4 \over 8} \,  {f(u)\over u}\,
\phi_k'(u)\, \phi_{-k}(u)\, \Biggl|_0^1\, .
\end{equation}
Following the prescription of  \cite{Son:2002sd} the 
retarded correlator is then computed as
\begin{equation}
G_{xy xy}(\omega,q)
 = - {\pi^2 N^2 T_0^4 \over 4} \, \lim_{\epsilon\rightarrow 0}
 \,  {f(\epsilon)\, \phi_k'(\epsilon)  \over \epsilon\, \phi_{k}(\epsilon)}\,.
\end{equation}
(The solution $\phi_k(u)$ has been normalized to
 $1$ at $u=\epsilon$.) We find
\begin{equation}
G_{xy xy}(\omega,q)
 = {i \pi N^2 T_0^3 \omega c \over 8} 
\, U(\kappa_1, \kappa_2, \kappa_3)\,.
\end{equation}
The Kubo formula  (\ref{kubo_visc}) gives the shear viscosity
\begin{equation}
\eta = {\pi\over 8} N^2 T_0^3 \, 
\sqrt{(1+\kappa_1)(1+\kappa_2)(1+\kappa_3) } = {\pi N^2 T_H^3
\, 
 \prod\limits_{i=1}^3 (1+\kappa_i)^{2}\over (2 +  \kappa_1 +  \kappa_2+ \kappa_3 - \kappa_1 \kappa_2 \kappa_3)^3}\,.
\label{visc_kubo_3}
\end{equation}
Comparing this result with the expression (\ref{entropy_density})
for the entropy density we immediately conclude that for any value of the chemical potential
\begin{equation}
{\eta\over s} = {1\over 4 \pi}\,.
\label{the_ratio}
\end{equation}
For small $\kappa_i$ we have 
\begin{equation}
\eta = {\pi N^2 T_H^3\over 8} \left( 1 + {1\over 2} 
\sum\limits_{i=1}^3 \kappa_i + O(\kappa_i^2)\right)\,. 
\end{equation}

\section{Shear viscosity from the diffusion pole}
\label{shear}

In this Section we explicitly compute the retarded correlation functions
of the stress-energy tensor and the R-currents and show that they exhibit 
a diffusion pole predicted by hydrodynamics. The value of the shear
 viscosity extracted from the pole is in agreement with 
Eq.~(\ref{visc_kubo_3}). For simplicity, in the rest of the paper 
we restrict ourselves to the case of a single charge black hole.
 We set $q_1 \neq 0$, $q_2=q_3=0$ and omit the 
index ``1'' in all subsequent expressions.

\subsection{The single charge black hole background} 

For a single charge black hole the effective Lagrangian 
(\ref{lagrangian}) can be written as
\begin{equation}
{{\cal L}\over \sqrt{-g}} =  
 R + {2\over L^2} {\cal V} -\frac{L^2}{8} H^{4/3} F^2
-\frac{1}{3} 
H^{-2} g^{\mu\nu} \del_\mu H\, \del_\nu H \,, 
\label{lagrangian_1}
\end{equation}  
where  $F_{\mu\nu}$ is the field-strength tensor of a $U(1)$ gauge
field, and ${\cal V}$ 
is the potential for the scalar field $H$,
\begin{equation}
{\cal V} = 2 H^{2/3}+4 H^{-1/3}\,.
\label{pot}
\end{equation}
The system 
of the  gauged supergravity equations of motion 
 for the fields 
$g_{\mu\nu}$, $A_{\mu}$, $H$ reads
\begin{equation}
\begin{split}
\,& \jsquare H = H^{-1} g^{\mu\nu} \del_\mu H \, \del_\nu H 
+\frac{L^2}{4} H^{7/3} F^2-\frac{3}{L^2} H^2 \frac{\del \calv}{\del H}\,,\\
\,&\del_\mu\left(\sqrt{-g} H^{4/3} F^{\mu\nu}\right) =0\,,\\
\,& R_{\mu\nu}=\frac{L^2}{4} H^{4/3} F_{\mu\gamma} F_{\nu}\ ^\gamma
+\ft 13 H^{-2} \del_\mu H \del_\nu H-g_{\mu\nu} \left[
\frac{2}{3 L^2} \calv+\frac{L^2}{24} H^{4/3} F^2\right]\,.
\end{split}
\label{eom}
\end{equation}
Consider small perturbations of the single charge background
\begin{eqnarray}
g_{\mu\nu} &=& g_{\mu\nu}^0 + h_{\mu\nu}(u,t,z)\,, \\
A_\mu &=& A_\mu^0 + A_{\mu}(u,t,z)\,,
\end{eqnarray}
where $g_{\mu\nu}^0$, $A_\mu^0$ are given by 
Eqs.~(\ref{metric_u_3}), (\ref{scal_gauge_u_3}) 
with $\kappa_1\equiv \kappa$, $\kappa_2=\kappa_3=0$.
We assume  that the  perturbations  depend on time, radial coordinate, 
and only one of the spatial world-volume coordinates, $z$.
We are interested in gravitational fluctuations 
 of the shear type, where 
the only nonzero components of $h_{\mu\nu}$ are $h_{t a}$, $h_{z a}$, $a=x,y$. 
One can show that fluctuations of all other fields except 
$A_a (r,t,z)$, $a=x,y$, can be consistently set to zero.
Introducing the new variables
\begin{equation}
T\equiv H_{ta} = g^{x x} h_{t a}\,,\;\;\;   Z \equiv H_{za} 
= g^{x x} h_{z a}\,,\;\;\; 
A = {2 A_a\over \mu}
\label{rescaling}
\end{equation}
the system of linearized equations derived from Eqs.~(\ref{eom}) can be written as
\begin{subequations}
\begin{eqnarray}
&& T' + {\qn \,  f \over \wn \, H } \; Z' + 
{\kappa  u \over 2 \, H}\;  A =0\,,\stru 
\label{eqq1r}\\
&& T'' + {u  H' -  H\over u H}\; T' - { \wn \, \qn \over f u} \; Z - 
 { \qn^2\over f u} \; T + {\kappa \, u\over 2\,  H} 
 A ' =0\,, \stru \label{eqq2r}\\
&& Z'' + {u f' - f \over u f} 
\; Z' +  { \wn^2 \, H\over  f^2 u}\; Z
+  { \qn \, \wn \, H\over  f^2 u} \; T =0\,, \stru \label{eqq3r}\\
&&
\left( H f A' + 2(1+ \kappa)\,  T \right)' -  { \qn^2 H\over u}\; A
+ { \wn^2 H^2\over  f u} \; A =0\,.\stru \label{eqq4r}
\end{eqnarray}
\end{subequations}
These equations are not independent: combining Eq.~(\ref{eqq1r})
with Eq.~(\ref{eqq2r}), one obtains Eq.~(\ref{eqq3r}). Thus it is sufficient to 
consider
 Eqs.~(\ref{eqq1r}), (\ref{eqq2r}),
(\ref{eqq4r}). Expressing  $Z(u)$ from 
 Eq.~(\ref{eqq2r}), we differentiate it  
with respect to the radial coordinate $u$
 and substitute the resulting expression for $Z'(u)$ into Eq.~(\ref{eqq1r}).
Thus we obtain  a
 system of two coupled differential equations for $G(u)\equiv T'(u)$
 and $A(u)$
\begin{eqnarray}
G'' &+& \left( {H'\over  H} +{f'\over f}\right) G'
+ \left( {H'\over u H^2} - {f'\over u f H} + {\wn^2 H\over u f^2} -
{\qn^2 \over u f} \right) G \nonumber \\&+&
{\kappa u\over 2 H} A'' + {\kappa u H f' +
 \kappa f (2 H- u H')\over 2 f H^2} A' 
+ {\kappa \wn^2\over 2 f^2} A =0\,,
\label{eq_G}
\end{eqnarray}
\begin{equation}
A'' + \left( {f'\over f} + {H'\over H}\right) \, A'
+ {\wn^2 H - \qn^2 f\over u f^2}\, A + {2 (1+\kappa)\over f H} \, G =0\,.
\label{eq_A}
\end{equation}
(For $\kappa=0$ Eqs.~(\ref{eq_G}), (\ref{eq_A})   decouple\footnote{ 
One should recall the rescaling (\ref{rescaling}).} and 
reduce respectively  to Eqs. (6.15) and (5.5d) of
  Ref.~\cite{Policastro:2002se}.)
For the system (\ref{eq_G}), (\ref{eq_A}), 
the exponents at the singular point $u=1$ corresponding to the horizon 
 are 
\begin{equation}
\alpha = \left\{\;  0\,, \; 0\,, \; i \wn U(\kappa)\,,  \; - i \wn U(\kappa)\;  \right\}\,,
\end{equation}
where $U(\kappa) \equiv U(\kappa,0,0)$.
Choosing the exponent $ - i \wn U(\kappa)$ corresponding to 
the incoming wave boundary condition at $u=1$, we 
look for the solutions to  Eqs.~(\ref{eq_G}), (\ref{eq_A})
in the hydrodynamic approximation in the form
\begin{eqnarray}
G(u) &=& - {u (2+\kappa u)\over 2 H(u)^2}\, f^{-i \wn U(\kappa)}
\left( G_0 (u) + \wn G_1 (u) + \qn^2 G_2 (u) + \cdots \right)\,,
\\
A(u) &=& {1\over \kappa H(u)}\,  f^{-i \wn U(\kappa)}
\left( A_0 (u) + \wn A_1 (u) + \qn^2 A_2 (u) + \cdots \right)\,,
\label{ans}
\end{eqnarray}
where functions $G_i$, $A_i$ are regular at $u=1$.
We obtain
\begin{eqnarray}
G_0(u) &=& C_1 - {C_2\over \kappa}\,, \\
A_0(u) &=& C_1 + u C_2\,, 
\end{eqnarray}
where $C_1$, $C_2$ are the integration constants. Next,
\begin{eqnarray}
G_1 &=& {i\over \kappa \, \sqrt{\kappa+1} \, (\kappa +2)\,  u\,
 (2+\kappa u)} 
\Biggl\{
(u-1)\, \Biggl( C_2 (\kappa +2)\, (2+\kappa u)\nonumber \\ &-& 
 C_1 \, \kappa \,  (2+\kappa +\kappa u+\kappa^2 u) \Biggr)
 \nonumber \\ &+& 2\, (\kappa+1)\, (C_2+u C_1)\, u \, (2+\kappa u) 
\log{{\kappa +2\over u + H(u)} }\Biggr\}\,,
\\
A_1 &=& - {i \over \sqrt{\kappa+1}\, (\kappa +2)} \Biggl[ 
(u-1)\,  \Biggl( C_1\, \kappa - C_2\, (\kappa+2) \Biggr) \nonumber \\
&+&
2\, (\kappa +1)\, \left( C_1 + C_2 u \right)
 \log{{\kappa +2\over u + H(u)} }\Biggr]\,.
\end{eqnarray}
Functions $G_2(u)$, $A_2(u)$ are given by rather cumbersome expressions
involving polylogarithms. We do not write them explicitly here.

The integration constants $C_1$, $C_2$ can be expressed 
in terms of the boundary values of the fields $T^{(0)}$, 
 $Z^{(0)}$, $A^{(0)}$ by solving the equations
\begin{eqnarray}
\lim_{u\rightarrow 0} A(u) &=& A^{(0)}\,,\label{lim1}\\
\lim_{u\rightarrow 0}
 u f(u)\left( G' - {1\over u H}\; G + {\kappa \, u\over 2\,  H(u)} 
 A ' \right) &=& \wn \qn\,  Z^{(0)} + \qn^2 \,T^{(0)} \,,
\label{lim2}
\end{eqnarray}
where Eq.~(\ref{lim2}) comes from Eq.~(\ref{eqq2r}).
We find
\begin{eqnarray}
C_1 &=& \kappa\,  A^{(0)}\,, \label{llim1}\\
C_2 &=&  -{2\kappa\, (1+\kappa)\, 
 \qn^2 \over \qn^2 - i\, 2 \, \sqrt{1+\kappa}\,  \wn}\,
 T^{(0)} + {\kappa^2\,  (\qn^2 - i  \, \sqrt{1+\kappa}\, \wn )
\over \qn^2 - i\,  2 \, \sqrt{1+\kappa} \, \wn} \,  A^{(0)}\,.
 \label{llim2}
\end{eqnarray}
One can observe the appearance of the hydrodynamic pole 
 in Eq.~(\ref{llim2}).

\subsection{The correlators}

We are interested in two-point retarded correlation functions 
of stress-energy tensors and  $R$-currents defined by 
\begin{equation}
  G_{\mu\nu \lambda\rho} (\omega, \q)
  = -i\!\int\!d^4x\,e^{-iq\cdot x}\,
  \theta(t) \< [T_{\mu\nu}(x),\, T_{\lambda\rho}(0)] \>\,.
\end{equation}
\begin{equation}
  G_{\mu\nu \lambda} (\omega, \q)
  = -i\!\int\!d^4x\,e^{-iq\cdot x}\,
  \theta(t) \< [T_{\mu\nu}(x),\, J_{\lambda}(0)] \>\,.
\end{equation}
\begin{equation}
  G_{\mu\nu} (\omega, \q)
  = -i\!\int\!d^4x\,e^{-iq\cdot x}\,
  \theta(t) \< [J_{\mu}(x),\, J_{\nu}(0)] \>\,.
\end{equation}
To compute the correlators we need to consider the 
boundary action.
On shell, the action reduces to the surface terms, 
$S= S_{horizon} + S_\epsilon$, where
\begin{eqnarray}
S_\epsilon &=& 
\lim_{u\rightarrow 0} {\pi^2 N^2 T_0^4\over 8} \int d^4 x\Biggl[
 -(1+\kappa )
-
{H\over u} H_{ta}' H_{ta} + {f\over u} H_{za}' H_{za}
 + {f H\over 2\pi^2 T_0^2} A_a' A_a
\nonumber \\
&-& {3\over 2} (1+\kappa)\,  H_{ta}^2 
- {1\over 2} (1+\kappa)\,  H_{za}^2 
+ {\sqrt{2 \kappa (1+\kappa)}\over 2\pi T_0} H_{ta} A_a\Biggr]\,.
\label{boundary_action}
\end{eqnarray}
The first term in Eq.~(\ref{boundary_action}), 
 $ -\pi^2 N^2 T_0^4 (1+\kappa) V_4/8$, is the density of the Gibbs potential 
$\Omega$  (i.e. the pressure with a minus sign), times  the four-volume.
The retarded two-point functions are obtained from $S_\epsilon$ 
following the recipe formulated in \cite{Son:2002sd}. After substituting 
the solution (\ref{ans}) into  Eq.~(\ref{boundary_action}), the
 part of the boundary action quadratic in fluctuations assumes the form
\begin{equation}
S_\epsilon^{(2)} = \int {d\omega d q\over (2\pi)^2}\, 
\phi_i^{(0)}(\omega, q)\,
  {\cal F}_{ik} (\omega,q)\, \phi_k^{(0)} (-\omega, -q)\,,
\end{equation}
where $\phi_i^{(0)}$ denote the boundary values of the fields 
$T^{(0)}$, $Z^{(0)}$, $A_a^{(0)}$. Then the retarded correlators 
are given by\footnote{Note that we obtain the correlators with upper indices.
Indices of the boundary theory correlators are raised or lowered with 
the flat Minkowski metric, so that e.g. $G^{t a z a} = - G_{t a z a}$.} 
\begin{equation}
   G^R   \;=\;
  \begin{cases}  - 2  {\cal F}_{ik} (\omega,q)\,,
                                                    & i=k , \\
           \noalign{\vskip 4pt}
           -  {\cal F}_{ik} (\omega,q)\,,
                                                    & i\neq k . 
         \end{cases}
   \label{f_fth}
\end{equation}
Computing the correlators, to leading order in 
 the hydrodynamic 
approximation we obtain\footnote{Contact terms are ignored.}
\begin{equation}
G_{t a t a} (\omega, q) = {\sqrt{1+\kappa} N^2 \pi T_0^3 q^2\over 8
(i \omega - {\cal D} q^2)}\,, 
\label{TX_TX_correlator}
\end{equation}
\begin{equation}
G_{t a z a} (\omega, q) = - {\sqrt{1+\kappa} N^2 \pi T_0^3 \omega q \over 8
(i \omega - {\cal D} q^2)}\,, 
\label{TX_ZX_correlator}
\end{equation}
\begin{equation}
G_{z a z a} (\omega, q) = {\sqrt{1+\kappa} N^2 \pi T_0^3 \omega^2\over 8
(i \omega - {\cal D} q^2)}\,, 
\label{ZX_ZX_correlator}
\end{equation}
where the diffusion constant is given by
\begin{equation}
 {\cal D} = {1\over 4\pi T_0 \sqrt{1+\kappa}} = {1\over 4\pi T_H}
 {1+\kappa/2\over 1+\kappa}\,.
\label{drel}
\end{equation}
In the limit $\kappa \rightarrow 0$ the results (\ref{TX_TX_correlator}) -
(\ref{drel}) reduce to those obtained in  
\cite{Policastro:2002se} for the case of a zero chemical potential.
For the correlators of the components
 of a stress-energy tensor and an $R$-current  we have 
\begin{equation}
G_{t a a} (\omega, q) =  {i \sqrt{2 \kappa (1+\kappa)} 
N^2 \pi T_0^3 \omega  \over
8 (i \omega - {\cal D} q^2)}\,.
\label{current_TX_correlator}
\end{equation}
\begin{equation}
G_{z a a} (\omega, q) = - {\sqrt{2 \kappa}\, 
N^2 T_0^2 \omega q \over
32 (i \omega - {\cal D} q^2)}\,.
\label{current_ZX_correlator}
\end{equation}
These correlators vanish in the limit $\kappa\rightarrow 0$, 
in agreement with \cite{Policastro:2002se}.
Finally, the retarded correlator of the $x$ (or $y$) component of the 
R-currents is given by
\begin{equation}
G_{x x} (\omega, q) = G_{y y} (\omega, q) = 
G_{a a} (\omega, q) = {i \kappa  N^2  T_0^2 \omega \over 16
(i \omega - {\cal D} q^2)} + O(\omega, q^2)\,.
\label{current_correlator}
\end{equation}
In the limit $\kappa\rightarrow 0$ the leading
 contribution in Eq.~(\ref{current_correlator}) vanishes. 
The subleading term gives $G_{a a} = -i N^2 T_0 \omega/16\pi$ which 
coincides with the result obtained in \cite{Policastro:2002se}.

In the limit of vanishing spatial momentum 
the nontrivial contribution to $G_{a a} (\omega, q)$ 
again comes from the subleading term 
in  Eq.~(\ref{current_correlator}). It is given by
\begin{equation}
G_{a a} (\omega, 0) = - {i (\kappa+2)^2  N^2 T_0  \omega \over 64 \pi
\sqrt{\kappa +1}}\,.
\label{current_correlator_q=0}
\end{equation}
We also find that in the limit  $q\rightarrow 0$
the correlators $G_{t a a}$ and  $G_{t a t a}$
vanish (modulo contact terms).

\subsection{The diffusion pole}
All the retarded correlators in the shear channel exhibit a diffusion pole
with the dispersion relation 
\begin{equation}
\omega = -i {\cal D} \, q^2 \,,
\label{drelm}
\end{equation}
where the diffusion constant $ {\cal D}$ is given by Eq.~(\ref{drel}).
To find viscosity, recall that in hydrodynamics 
$$
 {\cal D} = {\eta \over \varepsilon + P}\,.
$$
From thermodynamics it follows that 
$$
\varepsilon + P = T_H s + \mu \rho =  {2(1+\kappa)\over 2+\kappa}\; T_H s\,.
$$
Thus for the ratio of shear viscosity to entropy density we find
$\eta / s = 1 / 4\pi$
which coincides with the result (\ref{the_ratio}) 
obtained from the Kubo formula.

\section{Thermal conductivity}
\label{conductivity}

Thermal conductivity $\kappa_T$ can be computed using the appropriate Kubo formula
(see Appendix \ref{kubo_tc})
\begin{equation}
   \kappa_T = -\frac{(\epsilon+P)^2}{\rho^2 T}\lim_{\omega\to0}\;
   \frac1\omega \, \Im G(\omega, {\bf 0})\,.
\label{kubo_termal_conductivity}
\end{equation}
 Here $G$ is the retarded Green's function of the R-current components $J^x$
given by Eq.~(\ref{current_correlator_q=0}).
Thus we find
\begin{equation}
\kappa_T =  \frac{N^2 (\kappa+2)}{32 \pi}\; 
\frac{(\varepsilon +P)^2}{\rho^2} =  \pi N^2 T_H^2 \;  \frac{(1+\kappa)^2}{\kappa (\kappa +2)}\,.
\end{equation}
In terms of the chemical potential $\mn$ the thermal conductivity 
 can be written as
\begin{equation}
\kappa_T = \pi N^2 T_H^2 \; 
\frac{1+\sqrt{1-4 \mn^2}-\mn^2 (\sqrt{1-4 \mn^2}-5)}{8 \mn^2}\,.
\end{equation}
Comparing this result to the one for the shear viscosity
 (\ref{visc_kubo_3}), we observe that 
for a single-charge black hole, the shear viscosity 
and thermal conductivity can be expressed in terms of the 
chemical potential $\mu_1\equiv \mu$ as
\begin{equation}
\eta = {\pi N^2 T_H^3\over 8}\, F_\eta (\mu, T_H)\,, \qquad 
\kappa_T = {\pi N^2 T_H^2\over 8}\, F_\kappa (\mu, T_H)\,,
\end{equation}
where the functions  $F_\eta$, $F_\kappa$ depend only on the ratio
$\mn =  \mu/2\pi T_H$,
\begin{equation}
 F_\eta (\mu, T_H) = {8 \mn^2 \left( 1 - \sqrt{1 - 4 \mn^2}-\mn^2\right)^2
\over \left( 1 - \sqrt{1 - 4 \mn^2}\right)^3}\,,\qquad
 F_\kappa (\mu, T_H)  = {2\over \mn^2} \, F_\eta (\mu, T_H) \,.
\end{equation}
Thus for all values of $T_H$ and $\mu$ one finds an analogue of the 
Wiedemann-Franz law \cite{landau_10}
\begin{equation}
{\kappa_T \mu^2\over \eta T_H} = 8 \pi^2\,.
\end{equation}
For small $\mn$ we get
$$
F_\eta =  1 + \mn^2 - \mn^6 + O(\mn^8)\,.
$$
The function $F_\eta (\mn)$ is shown in Fig. 1.

\section{Critical behavior of
 transport coefficients}
\label{critical}

The boundary of thermodynamic stability is $\mn_c = 1/2$ or $\mu_c = \pi T$.
Expanding the shear viscosity and the thermal conductivity 
near $\mn =\mn_c$ we obtain
\begin{eqnarray}
\eta &=& \eta_* \left[ 1 + {2\over 3}\, \sqrt{\mn_c - \mn} -{20\over 9} 
\, (\mn_c-\mn) + O\left( (\mn_c-\mn)^{3/2}\right)
\right]\,,\\
\kappa_T &=& \kappa_{T *} \left[ 1 + {2\over 3}\, \sqrt{\mn_c - \mn} + {16\over 9} 
\, (\mn_c-\mn) + O\left( (\mn_c-\mn)^{3/2}\right)
\right]\,,
\end{eqnarray}
where
$$
\eta_* = {9 \pi N^2 T_H^3\over 64}\,, \qquad \kappa_{T *} =
 {9 \pi N^2 T_H^2\over 8}\,.
$$
Both the viscosity and the thermal conductivity are 
 finite\footnote{Curiously, the shear 
viscosity of 
He$\,^4$ near the $\lambda$-point is also finite and 
its first derivative is divergent, as first shown
theoretically by
 A.~M.~Polyakov \cite{polyakov} and later confirmed 
experimentally~\cite{biskeborn}.}
 at the critical point. Their derivatives diverge with
 the critical index equal to $1/2$.

\begin{FIGURE}[t]
{
  \parbox[c]{\textwidth}
  {
  \begin{center}
  \psfrag{X}{$ \mn$}
  \psfrag{Y}{$ F_\eta$}
  \includegraphics[width=3.8in]{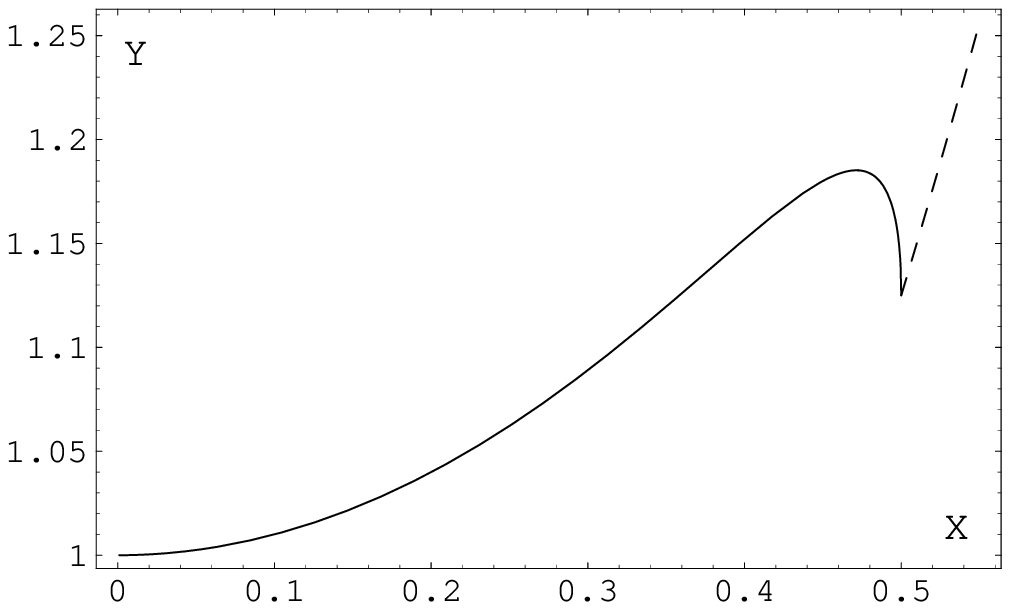}
  \caption
    {%
Normalized shear viscosity $F_\eta$ as a function of the chemical potential 
$\mn = \mu/2\pi T_H$. There is a cusp singularity at $\mu_c=\pi T_H$.
The part of the curve to the right of the singularity (shown in dashed line)
 is unphysical.
    }
  \end{center}
  }
\label{fig:complexplane-vector}
}
\end{FIGURE}

\section{Conclusion}
\label{conclusions}

We have considered the hydrodynamic regime of the ${\cal N}=4$
supersymmetric theory at finite temperature and finite chemical
potential.  We have computed the shear viscosity and the thermal
conductivity.  The shear viscosity is computed using two different
methods which give the same answer.  We find that the ratio of shear
viscosity and the entropy density is always equal to $1/4\pi$, which
is the same value found so far in all theories with gravity duals.
Our result demonstrates that the universality of this ratio extends to
theories with gravity duals at finite chemical potentials.  We found a
curious relationship between the shear viscosity and the thermal
conductivity similar to the Wiedemann-Franz law, and we have also
determined the critical behavior of the kinetic coefficients near the
boundary of thermodynamic stability.

One possible extension of this work is to compute the thermal
conductivity and diffusion coefficients in the case when all three
conserved charges are nonzero.  This would complete our knowledge of
the kinetic coefficents of ${\cal N}=4$ super-Yang-Mills theory in
the whole phase diagram, as the conformal invariance of the theory guarantees
 that the speed of sound is equal to $1/\sqrt{3}$ and the bulk viscosity 
 is zero for any temperature and chemical potential.

{\it Note added:} while this paper was being completed, we became aware of the work on 
the same subject by J.~Mas \cite{Mas:2006dy} 
whose results are in agreement with our analysis. 
Another recent paper on the subject is by K.~Maeda, M.~Natsuume, and T.~Okamura \cite{Maeda:2006by}.
A closely related work on the hydrodynamics of M2-branes by O.~Saremi appeared in \cite{Saremi:2006ep}.

\begin{acknowledgments}
 D.T.S. would like to thank the Perimeter
Institute and the Institute for Advanced Study, where part of this
work was completed, for hospitality.  The work of D.T.S. is supported,
in part, by DOE grant No.\ DE-FG02-00ER41132 and a grant-in-aid from
the IBM Einstein Endowed Fellowship.
 A.O.S. would like to thank A.~Buchel and R.~Myers for useful
discussions. Research at Perimeter Institute is supported in part by
funds from NSERC of Canada. 
\end{acknowledgments}

\appendix

\section{Rescaling the black hole solution}
\label{appendix_A}

The metric (\ref{metric_u_3})
 of the  gravity background dual to  ${\cal N}=4$ SYM with nonzero chemical potentials
 in Minkowski space can  be obtained by rescaling and taking an infinite volume limit
of the black hole solution with a spherical horizon. 
(At zero chemical potential, this procedure leads from the metric corresponding to an  
AdS-Schwarzschild black hole to the metric describing the near-horizon region of
 non-extremal three-branes.)
Following Refs.~\cite{Witten:1998zw}, \cite{Chamblin:1999tk}, here we explain
 how the rescaling works for the solution itself
 as well as for various thermodynamic quantities associated with it.

Focusing for simplicity on a single charge black hole, the metric 
(\ref{metric_k}) with $k=1$ reads
\begin{equation}
ds_5^2=-H^{-2/3} f_1 dt^2+ H^{1/3}\left(f_1^{-1} dr^2 +
r^2 \, d\Omega_{3}^2\right)\,,
\label{metric_bpz} 
\end{equation}
where
$$
f_1
=1-\frac{m_1}{ r^2}+  {r^2\over L^2}\,  H\,,\qquad H=1+\frac{q}{r^2}\,.
$$
Rescaling 
\begin{equation}
r\rightarrow \lambda^{1/4}\, r\,,\qquad  t\rightarrow \lambda^{-1/4}\, t\,, \qquad 
 m_1\rightarrow \lambda \, m_1\,, \qquad  q\rightarrow \lambda^{1/2} \, q\,, 
\end{equation}
and taking $\lambda \rightarrow \infty$ while simultaneously 
blowing up the sphere
\begin{equation}
 L^2\, d\Omega_{3}^2 \rightarrow \lambda^{-1/2}\,
 \left( dx^2 + dy^2 + dz^2\right)
\label{blow}
\end{equation}
in the limit we obtain the metric with a translationally invariant horizon
\begin{equation}
ds_5^2= - H^{-2/3} {r^2\over L^2} \, f dt^2+ H^{1/3}{L^2\over r^2  f}\, dr^2 
 +  H^{1/3} {r^2\over L^2}
\left( d x^2 + d y^2 + d z^2\right)\,,
\label{metric_bpzx} 
\end{equation}
where $f= 1+ q/r^2 - r_0^4/r^4$, with $r_0^4\equiv m_1 L^2$.

Similar reasoning applies to thermodynamic quantities and their densities.
Temperature, entropy, energy and the Gibbs potential
scale as
\begin{equation}
T_H\rightarrow \lambda^{1/4}\, T_H\,, 
\qquad S\rightarrow  S\,, \qquad E\rightarrow \lambda^{1/4}\, E\,, 
\qquad \Omega\rightarrow \lambda^{1/4}\,\Omega\,.
\end{equation}
For  the 
background (\ref{metric_bpz}), the inverse Hawking temperature, 
entropy, energy, and the thermodynamic potential $\Omega$
 are given correspondingly by (see e.g. \cite{Batrachenko:2004fd})
%
%
\begin{eqnarray}
&\,& \b = 2\pi\, L^2 \frac{(\rp^2+q)^{1/2}}{1+ q +2 \rp^2}\,,\qquad \qquad \;\;
S_{BH} = \frac{\pi^2}{2 G_5}\ \rp^2 
\left(\rp^2+q\right)^{1/2}\,,\nonumber \\
&\,& E = {\pi\over G_5} \left( {3 r_0^4\over 8 L^2} + {q\over 4}
+ {3 L^2\over 32}\right)\,, \qquad 
\Omega = {\pi\over G_5} \left(  - { r_0^4\over 8 L^2} + {r_+^2\over 4}
+ {3 L^2\over 32}\right)\,.\nonumber
\end{eqnarray}
In the limit $\lambda \rightarrow \infty$  for the rescaled quantities 
we find
\begin{eqnarray}
&\,& \b=2\pi\, L^2  \frac{(\rp^2+q)^{1/2}}{ q +2 \rp^2}\,,\qquad \qquad
S_{BH}= \lambda^{3/4}\, \frac{\pi^2}{2 G_5}\ \rp^2 
\left(\rp^2+q\right)^{1/2}\,,\nonumber \\
&\,& 
E = \lambda^{3/4} \, {\pi\over G_5}  {3 r_0^4\over 8 L^2}\,, \qquad \qquad  \qquad
\Omega =  - 
\lambda^{3/4} {\pi\over G_5}  { r_0^4\over 8 L^2}\,.
\end{eqnarray}
%
From (\ref{blow}) one can see that $2\pi^2L^3\rightarrow \lambda^{-3/4} V_3$
and thus the pressure and the densities of entropy and energy
are finite in the $\lambda \rightarrow \infty$ limit and are given by 
\begin{equation}
s= S_{BH}/V_3 = {r_+^2 (r_+^2+q)^{1/2}\over 4 G_5 L^3}\,,
\label{entrA}
\end{equation}
\begin{equation}
\varepsilon = E/V_3 = {3 r_0^4\over 16 \pi G_5 L^5}\,,
\label{energyA}
\end{equation}
\begin{equation}
P = -\Omega/V_3 = { r_0^4\over 16 \pi G_5 L^5}\,.
\label{PA}
\end{equation}
With  
identifications\footnote{Note that $r_0^4 = r_+^2 (r_+^2+q)$, 
where $r_+$ is the largest root of the equation $f(r)=0$.} (\ref{identif})
and $G_5= \pi L^3/2N^2$ from Eqs.~(\ref{entrA}), (\ref{energyA}), (\ref{PA})
   we obtain the expressions  (\ref{entropy_density}), (\ref{energy_density}),
 (\ref{pressure}) used in the main text.

\section{Relativistic hydrodynamics at finite chemical potential}
\label{kubo_tc}

For completeness here we review the hydrodynamics of a relativistic
fluid with one conserved charge.  The hydrodynamic equations include
the continuity equations
\begin{equation}
   \d_\mu T^{\mu\nu}=0, \qquad \d_\mu J^\mu =0
\end{equation}
and the constitutive equations, which formally have the form
\begin{equation}
   T^{\mu\nu} = (\epsilon+P) u^\mu u^\nu + P g^{\mu\nu}+\tau^{\mu\nu}
   ,\qquad
   J^\mu = \rho u^\mu + \nu^\mu
\end{equation}
Here $\epsilon$ and $P$ are the local energy density and pressure,
$u^\mu$ is the local velocity, $u_\mu u^\mu=-1$.  The parts
$\tau^{\mu\nu}$ and $\nu^\mu$ are the dissipative parts of the
stress-energy tensor and the current.  To complete the system of
equations we need expressions relating $\tau^{\mu\nu}$ and $\nu^\mu$
with derivatives of $u^\mu$ and of the thermodynamic potentials.

Following Landau and Lifshitz~\cite{Landafshitz6}, we can choose
$u^\mu$ and $\rho$ so that $\tau^{\mu\nu}$ and $\nu^\mu$ are
 orthogonal  to $u^\mu$
\begin{equation}\label{transverse}
   u_\mu \tau^{\mu\nu} = u_\mu \nu^\mu = 0\,.
\end{equation}
The most general form of the constitutive equation follows from the
second law of thermodynamics.  First we notice that
\begin{equation}\label{udT}
   u_\nu \d_\mu T^{\mu\nu} = - (\epsilon+P)\d_\mu u^\mu
   - u^\mu \d_\mu \epsilon + u_\nu \d_\mu \tau^{\mu\nu} =0
\end{equation}
Using the thermodynamic relations
\begin{equation}
   \epsilon+P = Ts + \mu \rho, \qquad d\epsilon = Tds + \mu d\rho,
\end{equation}
current conservation, and Eq.~(\ref{transverse}), Eq.~(\ref{udT}) can
be transformed into
\begin{equation}
   \d_\mu (su^\mu) = \frac\mu T \d_\mu \nu^\mu
    - \frac{\tau^{\mu\nu}}T \d_\mu u_\nu
\end{equation}
or
\begin{equation}
   \d_\mu \left(su^\mu - \frac\mu T\nu^\mu\right) =
   -\nu^\mu \d_\mu \frac\mu T - \frac{\tau^{\mu\nu}}T \d_\mu \nu^\mu\,.
\end{equation}
We now interpret the left hand side as the divergence of the entropy
current.  The right hand side thus must be positive.  This implies
\begin{align}
   \nu^\mu &= -\varkappa \left(\d^\mu\frac\mu T
   + u^\mu u^\lambda \d_\lambda \frac\mu T\right),\label{nu-constit}\\
   \tau^{\mu\nu} &= - \eta (\d^\mu u^\nu + \d^\nu u^\mu
   + u^\mu u^\lambda \d_\lambda u^\nu
   + u^\nu u^\lambda \d_\lambda u^\mu )
   -\left(\zeta-\frac23\eta\right)
   (g^{\mu\nu}+ u^\mu u^\nu) \d_\lambda u^\lambda\,.
\end{align}
Here $\eta$ and $\zeta$ are the shear and bulk viscosities,
respectively.  To have an interpretation of $\varkappa$ as the
coefficient of thermal conductivity, let us consider the case when
there is no charge transport, $J^i=0$, but there is an energy flow,
$T^{ti}\neq0$, which is the heat flow.  The local velocity $u^i$ is
necessarily small and is equal to
\begin{equation}
   u^i =  \frac{\varkappa}\rho \d^i \frac\mu T\,.
\end{equation}
Therefore
\begin{equation}
   T^{ti} = (\epsilon+P) u^i = \frac{\varkappa}\rho
   \d^i \frac\mu T  \, (\epsilon+P)\,.
\end{equation}
Using $dP=sdT+\rho d\mu$ one can write this equation as
\begin{equation}
   T^{ti} = - \varkappa \left( \frac{\epsilon+P}{\rho T}\right)^2
   \left( \d_i T - \frac T{\epsilon+P} \d_i P\right)\,.
\end{equation}
In the nonrelativistic theory the heat flow is proportional to the
gradient of temperature; in the relativistic limit there is an extra
contribution proportional to the gradient of pressure.  The
proportionality coefficient is the thermal conductivity,
\begin{equation}\label{kappavarkappa}
   \kappa_T =\left( \frac{\epsilon+P}{\rho T}\right)^2\varkappa\,.
\end{equation}

The Kubo's formula for $\kappa_T$ can be written down as follows.
Suppose one puts the thermal system in a slowly-varying external
background gauge field $A^\mu$ coupled to the conserved charge.  This
field will induce a current, proportional to the electric field $E_i =
\d_t A_i - \d_i A_t$.  But since $A_t$ plays the same role as the
chemical potential, by comparing with Eq.~(\ref{nu-constit}) we can
write
\begin{equation}
   J^i = \frac\varkappa T (\d^t A^i - \d^i A^t)
\end{equation}
In the case when the external fields are homogeneous in space, this
relation becomes very simple,
\begin{equation}
   J^i = i \frac\varkappa T \omega A^i\,.
\end{equation}
We can compare this relation with the one that follows from the linear response
theory, $J^i=-G_R(\omega,{\bf 0}) A^i$. Thus we obtain
\begin{equation}
   G_R (\omega, {\bf 0}) = -i \omega \frac\varkappa T\,.
\end{equation}
From Eq.~(\ref{kappavarkappa}) we find the Kubo formula (\ref{kubo_termal_conductivity}).


\begin{thebibliography}{99}



\bibitem{Maldacena:1997re}
  J.~M.~Maldacena,
   ``The large N limit of superconformal field theories and supergravity,''
  %
  Adv.\ Theor.\ Math.\ Phys.\  {\bf 2}, 231 (1998)
  [Int.\ J.\ Theor.\ Phys.\  {\bf 38}, 1113 (1999)]
  [arXiv:hep-th/9711200].

\bibitem{Gubser:1998bc}
  S.~S.~Gubser, I.~R.~Klebanov and A.~M.~Polyakov,
   ``Gauge theory correlators from non-critical string theory,''
  %
  Phys.\ Lett.\ B {\bf 428}, 105 (1998)
  [arXiv:hep-th/9802109].

\bibitem{Witten:1998qj}
  E.~Witten,
   ``Anti-de Sitter space and holography,''
  %
  Adv.\ Theor.\ Math.\ Phys.\  {\bf 2}, 253 (1998)
  [arXiv:hep-th/9802150].


\bibitem{Aharony:1999ti}
  O.~Aharony, S.~S.~Gubser, J.~M.~Maldacena, H.~Ooguri and Y.~Oz,
   ``Large N field theories, string theory and gravity,''
  %
  Phys.\ Rept.\  {\bf 323}, 183 (2000)
  [arXiv:hep-th/9905111].




\bibitem{Kovtun:2003wp}
  P.~Kovtun, D.~T.~Son and A.~O.~Starinets,
   ``Holography and hydrodynamics: Diffusion on stretched horizons,''
  %
  JHEP {\bf 0310}, 064 (2003)
  [arXiv:hep-th/0309213].

\bibitem{Kovtun:2004de}
  P.~Kovtun, D.~T.~Son and A.~O.~Starinets,
   ``Viscosity in strongly interacting quantum field theories from black hole
   physics,''
  %
  Phys.\ Rev.\ Lett.\  {\bf 94}, 111601 (2005)
  [arXiv:hep-th/0405231].




\bibitem{Buchel:2004qq}
  A.~Buchel,
   ``On universality of stress-energy tensor correlation functions in
   supergravity,''
  %
  Phys.\ Lett.\ B {\bf 609}, 392 (2005)
  [arXiv:hep-th/0408095].




\bibitem{Gubser:1998jb}
  S.~S.~Gubser,
  ``Thermodynamics of spinning D3-branes,''
  Nucl.\ Phys.\ B {\bf 551}, 667 (1999)
  [arXiv:hep-th/9810225].




\bibitem{Chamblin:1999tk}
  A.~Chamblin, R.~Emparan, C.~V.~Johnson and R.~C.~Myers,
  ``Charged AdS black holes and catastrophic holography,''
  Phys.\ Rev.\ D {\bf 60}, 064018 (1999)
  [arXiv:hep-th/9902170].




\bibitem{Cvetic:1999ne}
  M.~Cvetic and S.~S.~Gubser,
  ``Phases of R-charged black holes, spinning branes and strongly coupled
  gauge theories,''
  JHEP {\bf 9904}, 024 (1999)
  [arXiv:hep-th/9902195].



\bibitem{Behrndt:1998jd}
  K.~Behrndt, M.~Cvetic and W.~A.~Sabra,
  ``Non-extreme black holes of five dimensional N = 2 AdS supergravity,''
  Nucl.\ Phys.\ B {\bf 553}, 317 (1999)
  [arXiv:hep-th/9810227].



\bibitem{Cai:1998ji}
  R.~G.~Cai and K.~S.~Soh,
   ``Critical behavior in the rotating D-branes,''
  %
  Mod.\ Phys.\ Lett.\ A {\bf 14}, 1895 (1999)
  [arXiv:hep-th/9812121].


\bibitem{Gubser:2000ec}
  S.~S.~Gubser and I.~Mitra,
  ``Instability of charged black holes in anti-de Sitter space,''
  arXiv:hep-th/0009126.


\bibitem{Son:2002sd}
  D.~T.~Son and A.~O.~Starinets,
  ``Minkowski-space correlators in AdS/CFT correspondence: Recipe and
  applications,''
  JHEP {\bf 0209}, 042 (2002)
  [arXiv:hep-th/0205051].

\bibitem{Policastro:2002se}
G.~Policastro, D.~T.~Son and A.~O.~Starinets,
``From AdS/CFT correspondence to hydrodynamics,''
JHEP {\bf 0209}, 043 (2002)
[arXiv:hep-th/0205052].


\bibitem{Kovtun:2005ev}
  P.~K.~Kovtun and A.~O.~Starinets,
  ``Quasinormal modes and holography,''
  Phys.\ Rev.\ D {\bf 72}, 086009 (2005)
  [arXiv:hep-th/0506184].

\bibitem{Balasubramanian:1999re}
  V.~Balasubramanian and P.~Kraus,
   ``A stress tensor for anti-de Sitter gravity,''
  %
  Commun.\ Math.\ Phys.\  {\bf 208}, 413 (1999)
  [arXiv:hep-th/9902121].



\bibitem{Bianchi:2001kw}
  M.~Bianchi, D.~Z.~Freedman and K.~Skenderis,
  ``Holographic renormalization,''
  Nucl.\ Phys.\ B {\bf 631}, 159 (2002)
  [arXiv:hep-th/0112119].

\bibitem{Skenderis:2002wp}
  K.~Skenderis,
  ``Lecture notes on holographic renormalization,''
  Class.\ Quant.\ Grav.\  {\bf 19}, 5849 (2002)
  [arXiv:hep-th/0209067].


\bibitem{Buchel:2003re}
  A.~Buchel and L.~A.~Pando Zayas,
  ``Hagedorn vs. Hawking-Page transition in string theory,''
  Phys.\ Rev.\ D {\bf 68}, 066012 (2003)
  [arXiv:hep-th/0305179].




\bibitem{Batrachenko:2004fd}
  A.~Batrachenko, J.~T.~Liu, R.~McNees, W.~A.~Sabra and W.~Y.~Wen,
  ``Black hole mass and Hamilton-Jacobi counterterms,''
  JHEP {\bf 0505}, 034 (2005)
  [arXiv:hep-th/0408205].




\bibitem{Witten:1998zw}
  E.~Witten,
  Adv.\ Theor.\ Math.\ Phys.\  {\bf 2}, 505 (1998)
  [arXiv:hep-th/9803131].








\bibitem{Landafshitz6}
L.D.~Landau and E.M.~Lifshitz, {\em Fluid mechanics}, Pergamon Press,
New York, 1987, 2nd ed.

\bibitem{landau_10}
 E.M.~Lifshitz and L.~P.~Pitaevskii, {\em Physical kinetics}, Pergamon Press,
Oxford, 1981.



\bibitem{polyakov}
A.~M.~Polyakov, ``Nonequilibrium processes in the critical region'', 
Zh. Eksp. Teor. Fiz. {\bf 57}, 2144 (1969); Soviet Phys. JETP {\bf 30}, 1164 (1970).

\bibitem{biskeborn}
R.~Biskeborn and R.~W.~Guernsey, Jr., ``Critical exponents for the shear viscosity of 
$\,^4$He at $T_{\lambda}$'', Phys. \ Rev. \ Lett. {\bf 34}, 455 (1975).





\bibitem{Mas:2006dy}
  J.~Mas,
  ``Shear viscosity from R-charged AdS black holes,''
  arXiv:hep-th/0601144.



\bibitem{Maeda:2006by}
  K.~Maeda, M.~Natsuume and T.~Okamura,
  ``Viscosity of gauge theory plasma with a chemical potential from AdS/CFT,''
  arXiv:hep-th/0602010.



\bibitem{Saremi:2006ep}
  O.~Saremi,
  ``The Viscosity Bound Conjecture and Hydrodynamics of M2-Brane Theory at
  Finite Chemical Potential,''
  arXiv:hep-th/0601159.






\end{thebibliography}
\end{document}